\documentclass[conference]{IEEEtran}
\usepackage{cite}

\usepackage{bigstrut}

\ifCLASSINFOpdf
 \usepackage[pdftex]{graphicx}
\else
\fi
\usepackage[cmex10]{amsmath}
\usepackage{algorithmicx}

\usepackage{array}
\usepackage{url}

\newcolumntype{L}{>{\centering\arraybackslash}m{1.1cm}}

\begin{document}
%
\title{Generating Online Social Networks based on Socio-demographic Attributes}

\author{\IEEEauthorblockN{Muhammad Qasim Pasta}
\IEEEauthorblockA{Karachi Institute of Economics and Technology \\
Karachi, Pakistan\\
Email: mqpasta@pafkiet.edu.pk}

%

\vspace{-8pt}

\and
\IEEEauthorblockN{Faraz Zaidi}
\IEEEauthorblockA{University of Lausanne, \\ 
Lausanne, Switzerland and \\
Karachi Institute of Economics and Technology \\
Karachi, Pakistan \\
Email: faraz@pafkiet.edu.pk}

\and
\IEEEauthorblockN{C\'{e}line Rozenblat}
\IEEEauthorblockA{University of Lausanne \\
Lausanne, Switzerland \\
Email: celine.rozenblat@unil.ch }
}

\maketitle

\begin{abstract}

Recent years have seen tremendous growth of many online social networks such as Facebook, LinkedIn and MySpace. People connect to each other through these networks forming large social communities providing researchers rich datasets to understand, model and predict social interactions and behaviors. New contacts in these networks can be formed due to an individual's demographic attributes such as age group, gender, geographic location, or due to a network's structural dynamics such as triadic closure and preferential attachment, or a combination of both demographic and structural characteristics.

A number of network generation models have been proposed in the last decade to explain the structure, evolution and processes taking place in different types of networks, and notably social networks. Network generation models studied in the literature primarily consider structural properties, and in some cases an individual's demographic profile in the formation of new social contacts. These models do not present a mechanism to combine both structural and demographic characteristics for the formation of new links. In this paper, we propose a new network generation algorithm which incorporates both these characteristics to model network formation. We use different publicly available Facebook datasets as benchmarks to demonstrate the correctness of the proposed network generation model. The proposed model is flexible and thus can generate networks with varying demographic and structural properties. 

\end{abstract}


%
\IEEEpeerreviewmaketitle

\section{Introduction}

Past decade has seen an exponential growth in the usage of online social networks such as Facebook, LinkedIn and MySpace \cite{boyd07} with hundreds of millions of users connecting to these networks everyday. These networks represent acquaintance relationships among individuals created through mutual consent. The field of social network analysis and complex networks has profited from these networks as they provide rich datasets for researchers to investigate various hypotheses and conjectures related to social behavior and social dynamics in our society \cite{lievrouw02,garton97}.

The set of social implications where online social networks play an important role is very large. For instance, on Facebook, people can post information about an event, which can then be shared by other people in their network and thus propagate information in this manner very efficiently to a large audience \cite{iribarren11}. Other applications include online marketing \cite{trusov08}, spreading viruses \cite{fan11} and community formation \cite{kumar06b} which can be studied using analysis methods, dynamic processes, network metrics, visualization techniques and clustering algorithms on large realistic datasets.


Substantial research has been conducted in modeling social networks where the objective has been to develop algorithmic models that can mimic the structure and evolution of real world networks. More often than not, researchers have targeted structural characteristics such as high clustering coefficient, small geodesic distance, degree distribution following power-law, assortative mixing and presence of communities in these networks \cite{holme02,catanzaro04,kumpula09,xu09,badham10}. Apart from the structural characteristics, another aspect of these networks are the demographic characteristics of individuals that play an important role in the formation of new links. Demographic characteristics include attributes such as age group of an individual, gender, geographic location, professional activity sector, personal interests and hobbies \cite{preston01}. 



Most of the network generation models proposed in the literature do not consider these demographic characteristics in the network generation process. Some models have been proposed in the literature with the concept of social spaces, distances and similitude to refer to the homophilic properties of individuals but the details of these properties are often omitted in these papers \cite{wong06,badham10,almeida13}. They directly utilise distances drawn from some distribution to refer to how close two individuals are, which in turn determines the probability of link formation among individuals. The reason might be the unavailability of large datasets with demographic properties, which are still difficult to obtain due to proprietary data, terms of use, and data protection of personal data.

The study of these models has a number of practical and theoretical applications as they give us a better understanding of how networks are organized, how they evolve over time and how structural dynamics impact the overall network properties. There are a number of applied fields where these models are useful such as they help to generate large networks with desired structural properties. They are also useful for simulation studies to examine different network processes taking place such as epidemic spreading, influence mining and formation of community structures \cite{badham10,zaidi13b}. Another applied application area for these models is to test various network sampling methods \cite{kurant11} as these models can generate networks with different sizes and structural properties. We can then test the correctness of different sampling methods by verifying whether the structural properties are preserved in the extracted samples. These models also provide well defined algorithmic procedures to generate real world like networks from which, strong theoretical results can be derived such as studying the formation of giant component, average geodesic distances between nodes and the process of triadic closures \cite{rapoport57,bollobas03,li05a}. 

Our premise from these works is that both structural and demographic characteristics play a pivotal role in the development of new links and thus it is more fitting to consider a mechanism which takes into account the position of an individual in the network (structural properties) and its socio-demographic properties. This helps to rationalize link formation between two individuals in a network and incorporates the two major catalysts that introduce new links in any social network.


In this paper, we extend the network generation model proposed by the authors in \cite{pasta13b}. The model considers both structural as well as demographic characteristics to generate social networks and is based on two steps: initialization and construction. We extend the construction step from the model of \cite{pasta13b} by introducing a parameter to tune degree assortativity \footnote{We refer to the structural property of high degree nodes connecting to other high degree nodes and low degree nodes connecting to other low degree nodes in this paper as assortative mixing or degree assortativity \cite{newman02}.} which helps us to generate desired values of assortative, or disassortative mixing, further enhancing the capabilities of the existing model.

We use ten different publicly available datasets from the famous social networking website Facebook to validate the proposed model as we are able to reproduce networks with similar structural properties, given demographic attributes in the exact same  proportion as in the original dataset. Furthermore, we also perform an empirical analysis of the model to study how various parameters effect different properties of networks generated by the proposed model.

The rest of the paper is organized as follows: We discuss a number of articles that propose network generation models in section \ref{sec:related}. In section \ref{sec:proposedequation}, we formulate a general form to incorporate demographic as well as structural characteristics to determine similarity among two nodes, which in turn drives the connectivity of the whole network. In section \ref{sec:proposed}, we provide the details of the proposed model which consists of two steps, \textit{initialization} and \textit{construction}. Section \ref{sec:experimental} describes the experimental set-up and the datasets used for comparative analysis followed by the results and explanation in section \ref{sec:results}. We perform empirical analysis using different parameter configurations to demonstrate the spectrum and robustness of the proposed model in section \ref{sec:empirical}. Finally, we conclude in section \ref{sec:conclusion} giving possible future research directions.

\section{Related Work}\label{sec:related}
The discovery of small world and scale free networks has revolutionized the way we study networks. Among other networks, social networks also exhibit small world and scale free properties. Watts and Strogatz (WS) \cite{watts98} proposed a model to simulate the occurrence of triadic closures (clustering coefficient) and the small world effect (short geodesic distances) in networks. Starting from a regular lattice, random rewiring of links with a certain probability transforms a regular lattice into a network commonly known as small world network. Albert and Barabasi (BA) \cite{barabasi99} introduced a model based on preferential attachment to simulate how networks with degree distribution following power-law occur in real world. These networks are commonly known as scale free networks. Starting from a few nodes, new nodes are introduced in the network which connect to older nodes with a probability proportional to the existing connectivity of the nodes. Nodes with higher degree have a higher probability of forming new links, as a result, networks with skewed degree distributions are generated.

Most of the early works followed by these two ground breaking models revolved around the idea of having a unified model to generate both small world and scale free networks. For example, Holme and Kim \cite{holme02} proposed a modification to the BA model adding a triad formation step after the preferential attachment step to create triads in a network. This step increases the overall clustering coefficient of a BA network, generating a network with both small world and scale free properties. Other variants of the BA model such as \cite{dorogovtsev02,klemm02,guo05,fu06,wang08b,li12a} produce networks having high clustering coefficient by introducing triads one way or the other and nodes connect using the preferential attachment rule to have a scale free degree distribution. Short average geodesic distances are not explicitly enforced but occur as a by-product of the scale-free behavior where most of the nodes are connected to a few nodes with very high degree, thus creating short average path lengths in the entire network through these high degree nodes, a phenomena termed as funnelling \cite{newman01b}.

Different researchers have used the idea of n-partite, and specially bi-partite graphs to generate social networks. The authors \cite{newman02a} introduce the idea to generate affiliation networks similar to co-authorship networks \cite{newman01} using random bipartite graphs with arbitrary degree distributions. This idea is also used by Guillaume and Latapy~\cite{guillaume04} as they identify bipartite graph structure as a fundamental model of complex networks by giving real world examples. The authors call the two disjoint sets of a bipartite graph as \textit{bottom} and \textit{top}. At each step, a new \textit{top} node is added and its degree \textit{d} is sampled from a prescribed distribution. For each of the \textit{d} edges of the new vertex, either a new \textit{bottom} vertex is added or one is picked among the pre-existing ones using preferential attachment. The bipartite graph is then projected as a unipartite graph to obtain a small world and scale free network. A more generalized model based on similar principles was proposed \cite{bu07} where instead of using the bipartite structure, a network can contain \textit{t} disjoint sets (instead of just two sets, as is the case of the bipartite graph). The authors discuss the example of sexual web~\cite{lilijeros01} which is based on the bipartite structure. A sexual web is a network where nodes represent men and women having relationships to opposite sex, and similar nodes do not interact with each other. At each time step, a new node and \textit{m} new edges are added to the network with the sum of the probabilities equal to $1$. The preferential attachment rule is followed as the new node links with the existing nodes with a probability proportional to the degree of the nodes. 

A growing network model \cite{catanzaro04} was proposed to incorporate the assortative mixing behavior in social networks. Assortative mixing here, refers to the structural property of individuals to connect with individuals having similar number of links. This model allows links to be added between existing individuals as well as new individuals on the basis of their degree thus forcing links between similar degree nodes, and inducing high assortativity in the network. 

Models based on demographic attributes have also been proposed where the goal is to determine connectivity based on social attributes. The social similarity, in these artefacts is often referred to as the social distance and the approach in general is termed as spatial approach for network generation. One such model based on social distance between individuals was presented by \cite{boguna04} where the model aims to generate networks with high clustering coefficient, assortativity and hierarchical community structures. Social distance refers to the degree of closeness or acceptance that an individual feels towards another individual in a social network. The closer two individuals are, the higher they have a probability to form a new link. The authors used a real acquaintance network to demonstrate the correctness of the proposed algorithm. Another model \cite{wong06} was proposed which uses spatial distance to model nodal properties and homophilic similarity among individuals. The model randomly spreads nodes in a geographical space such that the edge formation probability is dependent on the spatial distances among nodes. The network thus generated exhibits high clustering coefficient, small geodesic distance, power-law degree distribution, and the presence of community structures. 

A three phase spatial approach \cite{badham10} was proposed to generate networks with controllable structural parameters. This approach controls three important structural characteristics, the clustering coefficient, assortativity and degree distribution using input parameters making it quite useful to generate large networks. The model also takes as input, the degree sequence required in the final network. This static model uses a notational space to identify nodes closer to each other, a layout modification step to move nodes with similar degree closer and edge creation among nodes based on these spatial and layout modification step to achieve desired clustering coefficient and assortativity.

A very recent model focuses on the homophilic (referred to as demographic characteristics earlier in this paper) property of social networks \cite{almeida13}. The authors modify the BA model by introducing a homophilic term which creates regions where characteristics of individuals impact the rate of gaining links as well as links between individuals with similar and dissimilar characteristics. The authors introduce the notion of \textit{similitude} which represents the similarity of two individuals based on their intrinsic characteristics like children of the same
ethnicity are more likely to become friends in school and proteins with similar functions have a higher chance to connect to each other. As a result new connections are established by considering the degree and similitude of nodes but details of how this similitude can be calculated are not provided. The model maintains five important structural features, power-law degree distribution, preferential attachment, short geodesic distance, high clustering coefficient and growth over time. 

Evolutionary network models with aging nodes have also been proposed in the literature such as \cite{dorogovtsev00a,zhu03,geng09,wen11}. For example \cite{wen11}, the authors study the dynamic behavior of weighted local-world evolving networks with aging nodes. Newly added nodes connect to existing nodes based on a strength-age preferential attachment and the results show that the network thus generated has power-law degree distribution, high clustering coefficient and small world properties.

There exists a number of models based on the local-world phenomena \cite{pan06,sun07,wang09,wen11} where nodes only consider there neighbourhood in contrast to traditional network models that assume the presence of global information. For example \cite{wang09} investigate a local preferential attachment model to generate hierarchical networks with tunable degree distribution, ranging from exponential to power-law.

Another class of graphs models, the exponential random graph models have gained a lot of popularity \cite{frank86,snijders06,robins07} also known as $\mathit{p^*}$. These models are used to test, to what extent nodal attributes and structural dependencies describe structure of a network measured using frequency of degree distribution, traids and geodesic distances \cite{toivonen09}. The possible ties among individuals are modelled as random variables, and assumptions about dependencies among these random tie variables determine the general form of the exponential random graph model for the network \cite{robins07}. An important difference between network generation models and ERGMs is that network models try to explain how a network evolves whereas ERGMs do not explicitly explain any network generation process \cite{toivonen09}.

Models to generate clustered graphs also exist in the literature where the goal is to have community structures embedded in the resulting networks \cite{condon99, lancichinetti09, moriano13, zaidi13a,pasta13a}. For instance, \cite{pasta13a} recently proposed a tunable network generation model to ensure three structural properties of the communities embedded in the generated network, namely the connectivity within each community follows power-law, communities have high clustering coefficient and are organized in hierarchical structures. Since we do not address the issue of having community structures in the current work, the readers can go through \cite{pasta13a} to explore the literature related to network generation models with community structures.

An exhaustive review of network generation models is out of scope in this text, yet we have tried to cite a wide spectrum of different network generation models. Partial surveys, reports and comparative analysis for different network generation models can be found in  \cite{dorogovtsev02,newman03,jackson05,fortunato10,badham10,toivonen09,goldenberg10,zaidi13b}. None of the models to generate networks considersboth demographic and structural attributes during the network generation process at the same time. Our contribution lies in considering demographic as well as structural characteristics as the driving force for link formation between individuals. We provide exact details of calculating a similarity based on demographic properties and its utilization in link formation among individuals. The results we obtained from simulations using the proposed model demonstrate that the networks obtained are similar to different Facebook datasets in terms of geodesic distances, clustering coefficients, assortative mixing, density and degree distributions.

\section{Demographic and Structural Characteristics} \label{sec:proposedequation}

The proposed model is quite generic and aims to provide a general form to calculate similarity between any two individuals based on their demographic and structural characteristics. This general formulation can be further refined by adding more network specific details. First we introduce the general form, and then we provide details for the implementation of the model.

The premise upon which the proposed formulation is developed is that, for individuals $i$ and $j$, the link formation is a function $\mathit{f}(i,j)$ of two types of characteristics, demographic ($\mathit{D}$) and structural ($\mathit{S}$). Mathematically we can represent this relation as:

\begin{equation}
\mathit{f}(i,j)= \alpha \{ \mathit{D}_{i,j} \} + \beta \{ \mathit{S}_{i,j} \} 
\label{eq:f}
\end{equation}

$\mathit{D}_{i,j}$ and $\mathit{S}_{i,j}$ represent the demographic and structural similarities between $i$ and $j$ respectively and, $\alpha$ and $\beta$ represent equilibrium factors to control the balance between demographic and structural characteristics. Within this basic framework, different demographic and structural attributes can be considered. Specially for demographic characteristics, we propose a method to handle categorical, ordinal and numerical attributes separately, which allows us to incorporate any type of demographic attribute in the calculation of similarities between individuals, which in turn drives the network generation process. We discuss the calculation details below:


\subsection{Demographic Characteristics}\label{sec:demographic}

As discussed above, we consider different categorical, ordinal and numerical characteristics as demographic characteristics of an individual. For every categorical attribute $\mathit{C_p}$ where $p$ represents different attributes, the similarity between individuals $i$ and $j$ is assigned using the following equation:

\begin{equation}
\mathit{C_p}(i,j)=
\begin{cases}
0, \text{if $i_p=j_p$ }
\\
1, \text{if $i_p \neq j_p$ }
\end{cases}
\end{equation}

Similarly for every ordinal attribute $\mathit{O_q}$ where $q$ represents different attributes, the similarity between $i$ and $j$ is calculated using:

\begin{equation}
\mathit{O_q}(i,j)= \dfrac{|i_q - j_q|}{\mathit{\rho_q}} 
\end{equation}

where $i_q,j_q$ are the ranking orders, $|*|$ represents absolute value and $\mathit{O_q}$ is normalized using the maximum different ordinal values possible for attribute $q$ denoted by $\mathit{\rho_q}$ in the above equation. Similar to ordinal attributes, we calculate the normalized difference between numerical attributes of $i$ and $j$ using the following equation: 

\begin{equation}
\mathit{N_r}(i,j)= \dfrac{|i_r - j_r|}{\mathit{\rho_r}} 
\end{equation}

Low values of $\mathit{C_p}$,$\mathit{O_q}$ and $\mathit{N_r}$ suggest high similarity among individuals and high values indicated greater demographic distances among the individuals. Using the above equations, we can calculate an accumulative similarity value using equations 1, 2 and 3, based on demographic characteristics as follows:

\begin{equation}
\mathit{D}_{i,j}= \sum \mathit{C_p} + \sum \mathit{O_q} + \sum  \mathit{N_r}
\label{eq:d}
\end{equation}

The above equation shows a linear combination of a categorical, an ordinal and a numerical characteristic to give a general form where any number of such demographic attributes can be combined together.

\subsection{Structural Characteristics}\label{sec:structural}

In case of structural properties, we consider Preferential Attachment based on existing degree of nodes. For a newly added node $j$ (which initially will have zero connections), the probability of connecting to a node $i$ already existing in the network is directly proportional to the normalized degree of node $i$. The degree is normalized using the maximum node degree in the current network represented by $max(deg_n)$ as shown below:

\begin{equation}
\mathit{PA}(i,j) = 1- \dfrac{deg_i}{max(deg_n)} 
\label{eq:pa}
\end{equation}

We normalize this factor just to control the weight of each structural characteristic as all our characteristics are normalized between values 0 and 1. The values are subtracted from 1 to have low values represent similarity and high values represent dissimilarity among two nodes. 

We also consider the property of the triadic closures (commonly known as friend-of-a-friend phenomena in sociology) which gives us a high similarity if two nodes are connected by a common node. Preference for formation of triadic closures as $i$ and $j$ have common friends is calculated using the following equation:

\begin{equation}
\mathit{FoF}(i,j)= 1- \dfrac{ <i> {\cap} <j> }  { min (|i|,|j|) } 
\end{equation}

where $<i> {\cap} <j>$ represents the common friends of i and j and $ min (|i|,|j|) $ represents the minimum number of friends of either i or j. The minimum value in the denominator ensures that a relationship is not penalized just because one of the individual has high number of links. The more friends two individuals have in common, the more chances they have of forming a new link among themselves. Again the values are subtracted from 1 to ensure low values represent similarity.

\begin{equation}
\mathit{S}_{i,j}=  \mathit{PA} + \mathit{FoF}
\label{eq:s}
\end{equation}

Finally combining equation \ref{eq:d} and \ref{eq:s} as input to equation \ref{eq:f}, we can calculate an accumulated similarity for link formation between two individuals where both demographic as well as structural attributes are taken into account. Collectively, we refer to demographic and structural attributes as similarity based link formation.

All the above discussed demographic and structural attributed can be assigned a weight to associate more importance to any single property. The current experiments resulted in high similarity with the real data sets so we didn't explore this option further, but this remains an interesting path to be explored in the future.

\section{Proposed Model} \label{sec:proposed}

Apart from the distribution of demographic attributes, the model takes as input, the desired number of nodes in the network $n$, the minimum and maximum node degree $m_o$ and $ m_f$, the probability of triad formation $P(T_f)$, the probability of triad linkage $P(T_L)$, links added in linkage count $L$, assortativity parameter $(\gamma)$ and a similarity threshold ($th$). We also take weights $\omega$ for each demographic and structural attribute which can eventually help us to tune each characteristic's role in the formation of links among individuals. 

The parameters minimum and maximum node degree allow us to control the overall average node degree in the network. Probability of Triad formation determines whether a newly added node will form a triad with one of the neighbors of the node it connects to, based on similarity calculated in equation \ref{eq:s}. This helps to increase the clustering coefficient of the network. Triad linkage determines if new edges will be added as triadic closures to previously added nodes and linkage count determines how many such edges will be added. Again this further increases the overall clustering coefficient of the network. Assortativity parameter controls the assortative or disassortative mixing among nodes based on their degree. Finally the similarity threshold helps to enforce demographic homophily in the network as nodes connect to demographically similar nodes for low values of this parameter.

\begin{figure}
\begin{center}
\includegraphics[width=0.49\textwidth]{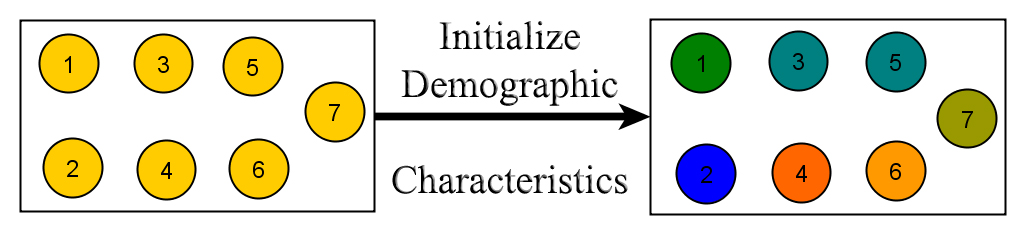}
\end{center}
\caption{The initialization step where nodes are randomly assigned demographic characteristics. Nodes are colored just to facilitate the readers to develop a mental map of the process. This color is assigned according to a combination of different characteristics where similar color represents similarity of nodes.}
\label{fig::initialization}
\end{figure}
\begin{figure*}
\begin{center}
\includegraphics[width=0.79\textwidth]{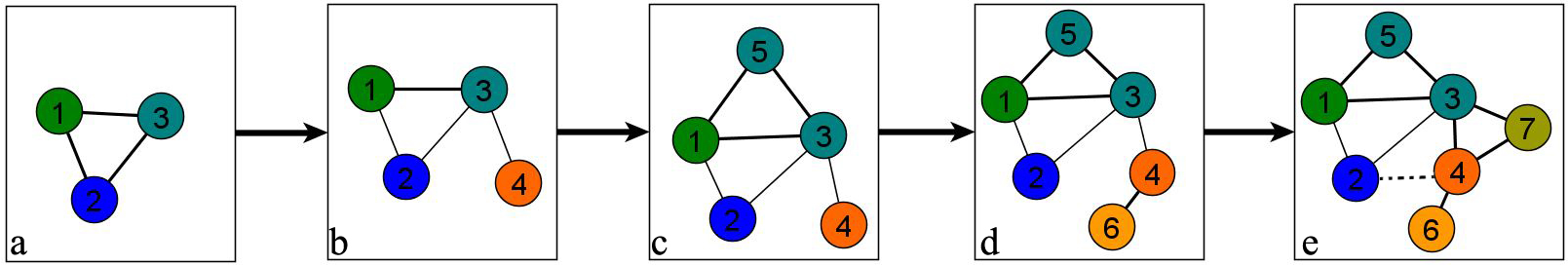}
\end{center}
\caption{Construction steps from (a) to (e) where initialized nodes are linked together based on demographic and structural characteristics. Step (a) shows a three nodes are added as a triad. Step (b) shows a node $4$ is added and the triad formation step and triad linkage steps are probabilistically not performed. Step (c) shows node $5$ is added connecting it to node $3$ based on similarity and a triad formation step connects it to node $1$. Step (d) where node $6$ is added. Step (e) where node $7$ is added with triad formation, and a new edge is created between node $4$ and $2$ based on triad linkage.}
\label{fig::construction}
\end{figure*}

The model comprises of two basic steps, the initialization step and the construction step. Within the construction step, three steps are performed, similarity based linking, triad formation, and linkage formation. The general idea is that a node first chooses another node to connect to, based on structural and demographic similarity (similarity based linking), connects probabilistically to one of its neighbors to form a triad, and then connects to some other nodes based on the principle of triadic closure. All these steps are described below:


\begin{enumerate}
\item \label{initialize} The initialization step randomly assigns demographic attributes in the given proportion to each of the $n$ nodes of the network. This results in a set of initialized nodes as shown in figure \ref{fig::initialization}. The nodes are numbered to associate a logical order which can be assigned randomly as the model is independent of this ordering of nodes.

\item \label{step1construction} To start construction of the connected network, the algorithm selects the first three nodes and connects them as a triad, irrespective of their similarity, as shown in figure \ref{fig::construction}(a).

\item \label{step2construction} A new node $n$ is then selected from the set of initialized nodes. A random number $m$ is generated between $m_o$ and $m_f$ to determine the number edges to be added to the network, with and without node $n$. While the added links in this iteration are less than $m$, the following three steps are repeated: 

\begin{enumerate}

\item \label{similaritybsed} An edge is created with node $n$ and an existing node $v$ with probability proportional to its similarity with $n$ i.e.\ the probability to create an edge between node $n$ and $v$ is given by:

\begin{equation}
\mathit{ P_{sim}(i) = \frac{f(i,j)}{\sum_{v\in\Gamma}f(i,v)}}
\label{eq:psim}
\end{equation}

where $\mathit{f(i,j)}$ is calculated as per equation \ref{eq:f} and Gamma ($\Gamma$) is a set of nodes that have a similarity from node $n$ greater than the similarity threshold ($th$). If the input parameter $th$ = 0.5, then we consider only nodes whose $f(i,j)>0.5$ to be able to connect to node $n$. This ensures that the demographic and structural homophily of the network is maintained and nodes connect to other similar nodes only. If no such node is found, $n$ is left disconnected. The effects of the parameter $th$ on the network structure are described in section \ref{sec:empirical}.


\item \label{triadformation}Based on the probability of triad formation $P(T)$, $n$ is then linked to a neighbor of the node it connected to in the previous step with probability $P_{t}$ according to probability given below:

\begin{equation}
\mathit{P_{i} =  \frac{deg_{i}^\gamma}{\sum_{v\in\tau}{deg_{v}^\gamma}}}
\label{eq:assor}
\end{equation}

where $deg_i$ represents the degree of node $i$ and $\tau$ represents the neighbourhood of node $i$. If no such node is found, or a link with such a node already exist, no new link is formed in this step. The $\gamma$ in the equation above incorporates assortative mixing which implies that individuals tend to connect to nodes with similar node connectivity, we use the method inspired from the work of \cite{guo06} where they have demonstrated that the introduction of parameter $\gamma$ with preferential attachment based on degree can generate assortative or disassortative networks. The effects of this parameter are demonstrated in section \ref{sec:empirical}.

\item \label{linkformation} Based on Triad linkage, a node which has already acquired some links (initially, it would be one of the three nodes that we connected in a triad) is selected randomly from the network and two of its neighbors connected by creating an edge between them. If an edge already exists between the two nodes, another node is selected to do the same. This process is repeated $L$ times to form $L$ triads in network. Essentially this step controls the overall clustering coefficient of the network and is used to tune the number of triads present in the network.

\item \label{edges}	The process is repeated from step \ref{similaritybsed} until the number of edges created in this iteration reaches $m$. Some edges are created for the newly added node $n$ whereas some are created among previously processed nodes. If no new edges can be created in any of the above three steps, the algorithm proceeds to step \ref{step2construction}. 
\end{enumerate}

\item \label{repeat} The process is repeated from step \ref{step2construction} until all the nodes in the initialized set are processed.

\end{enumerate}

The algorithm in a nutshell, adds a new node from the initialized step to the construction step and tries to add edges in three different ways: adds a new edge based on its similarity to existing nodes, possibly perform a triad formation step to create triads, and finally add links probabilistically between previously existing nodes to create more triads.

For clarification, we consider a small example with seven nodes. We consider the case of three demographic attributes, school (categorical), major (categorical) and age (numerical). Given as input, there are 3 possible schools in the proportion (2:2:3), there are two possible majors in the proportion (3:4) and the students have 3 possible age values in the proportion (3:3:1).
These attributes are assigned randomly to all the seven nodes as shown in figure \ref{fig::initialization} where the color coding in the initialization set depicts a unique color for a combination of attributes. So for nodes 3 and 5, the same color means that these individuals have exactly the same values for all demographic characteristics.

During the construction step, nodes from the initialization step are iteratively added to the network as shown in figure \ref{fig::construction}. Step (a) in figure \ref{fig::construction} shows that nodes 1, 2 and 3 are connected as a triad. Step (b) shows that node 4 is added to the network and connects to node 3 based on node similarity. Subsequently nodes 5,6 and 7 are added to the network where similar nodes form links on the basis of equation \ref{eq:f} and triad formation step introduces traids in the network.

\section{Data sets and Experimental Setup} \label{sec:experimental}
We used Facebook datasets publicly made available by \cite{traud11} representing the friendship social structure of different American colleges and universities at a single point in time. The demographic attributes present in the dataset are gender, class year, major and residence (housing). We used ten randomly chosen networks which are Caltech (769 nodes), Reed (962 nodes), Haverford (1446 nodes), Simmons (1518 nodes), Swarthmore (1659 nodes), Hamilton (2314 nodes), Oberlin (2920 nodes), Middlebury (3075 nodes), Wesleyan(3593 nodes) and American (6386 nodes). 

We tested our model to simulate networks of exactly the same size as that of these $10$ networks. The parameter configurations required for the proposed model to generate these networks are given in \ref{tbl:parameters}. We perform structural comparison between the original and the generated networks using density, geodesic distances, clustering coefficient, assortativity and power-law. We calculate the power-law fit using the method proposed by \cite{clauset09} and the assortative mixing value using \cite{newman03a}.

\begin{table}
  \centering
    \begin{tabular}{|l|L|L|L|L|L|} 
    \hline
    \textbf{Dataset} & \textbf{Minimum Number of Edges $m_o$} & \textbf{Maximum Number of Edges $m_f$} & \textbf{Probability of Triad Formation $P(T_f)$} & \textbf{Probability of Triad Linkage $P(T_L)$} & \textbf{Number of Links added in Triad Linkage $L$} \\
	\hline
    Caltech & 2     & 43    & 1     & 1     & 1 \\
    \hline
    Reed  & 2     & 43    & 0.5   & 0.5   & 1 \\
    \hline
    Haverford & 10    & 80    & 0.6   & 0.7   & 1 \\
    \hline
    Simmons & 2     & 43    & 0.9   & 0.7   & 2 \\
    \hline
    Swarthmore & 2     & 72    & 0.9   & 0.6   & 2 \\
    \hline
    Hamilton & 2     & 82    & 0.9   & 1     & 1 \\
    \hline
    Oberlin & 2     & 61    & 0.5   & 0.7   & 1 \\
    \hline
    Middlebury & 2     & 81    & 0.9   & 0.9   & 1 \\
    \hline
    Wesleyan & 2     & 77    & 0.9   & 0.6   & 2 \\
    \hline
    American & 2     & 68    & 0.9   & 0.7   & 2 \\
    \hline
    \end{tabular}%
	\vspace{5pt}
	\caption{Parameters used to generate graphs equivalent to original datasets from Facebook. The weight of Demographic and Structural Attributes were kept constant at 1 giving equal importance. Two other parameters, Assortative Mixing and Threshold were kept 0 and 0.5 respectively.} 
	\label{tbl:parameters}
\end{table}%



\section{Results and Discussion}\label{sec:results}

\begin{figure}
\begin{center}
\includegraphics[width=0.45\textwidth]{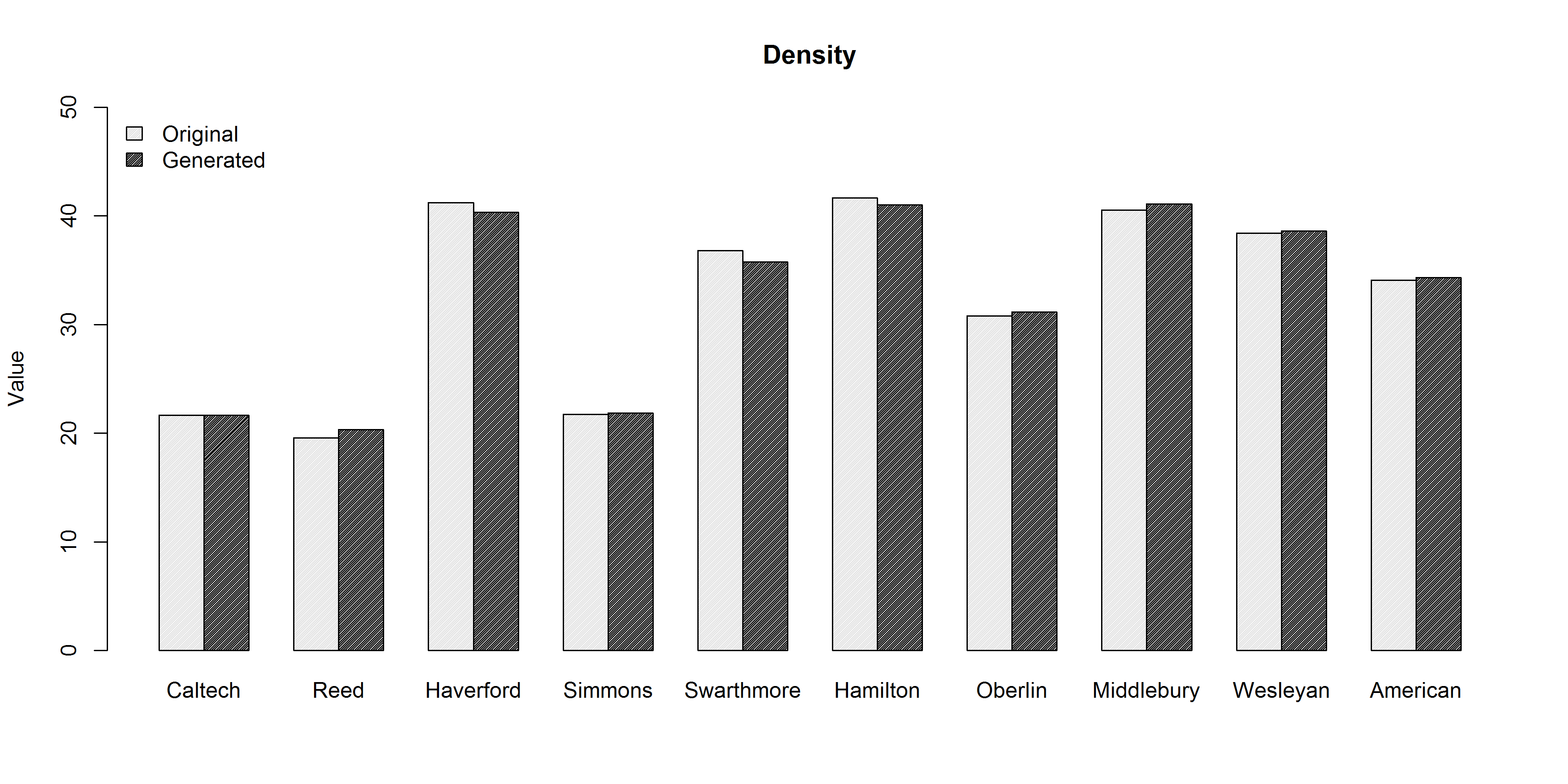}
\end{center}
\vspace{-10pt}
\caption{Comparative analysis of node-edge ratio or density of the original graphs and the generated graphs.}
\label{fig::density}
\end{figure}

\begin{figure}
\begin{center}
\includegraphics[width=0.45\textwidth]{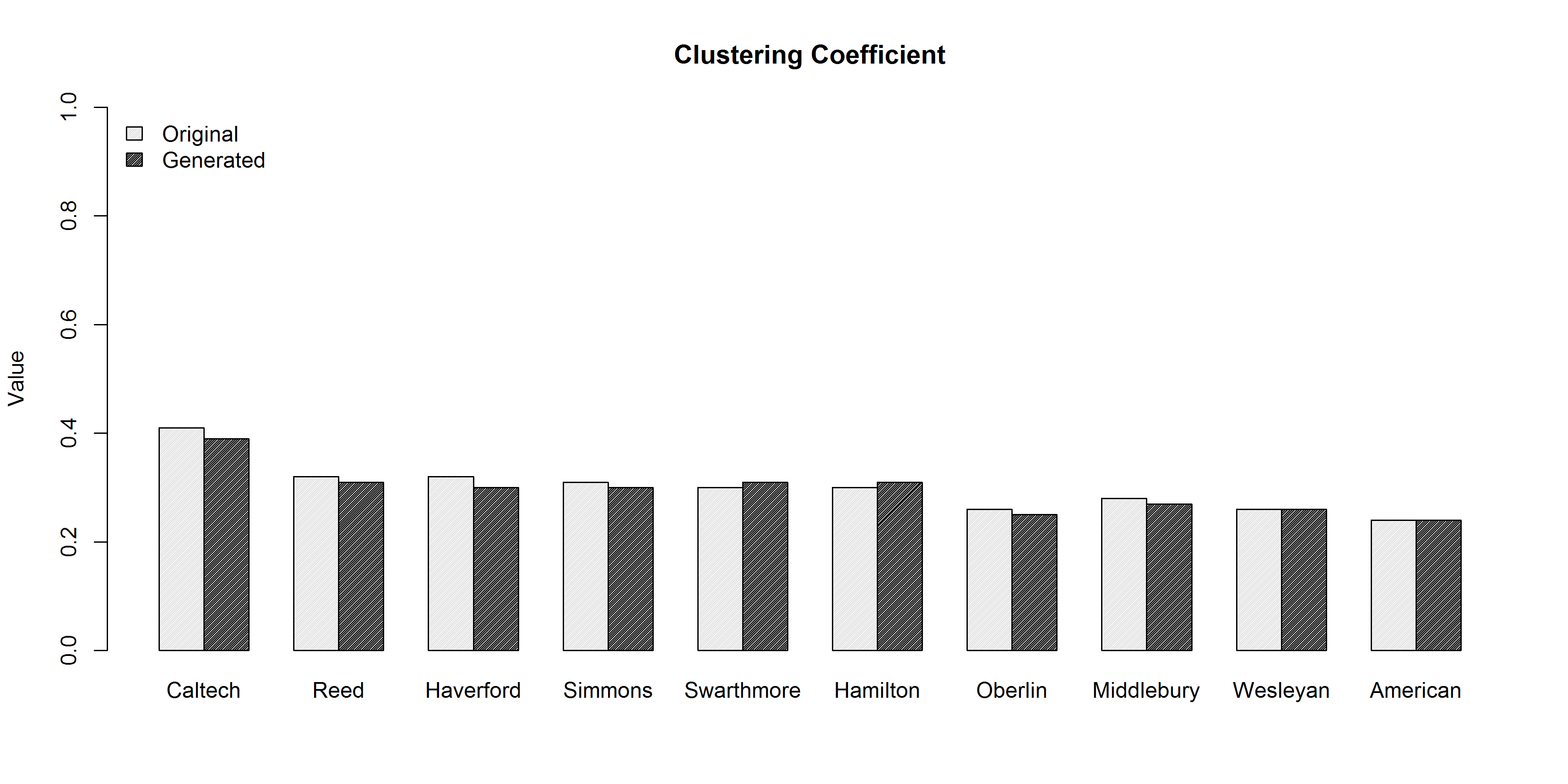}
\end{center}
\vspace{-10pt}
\caption{Comparative analysis of clustering coefficient of the original graphs and the generated graphs.}
\label{fig::cc}
\end{figure}

\begin{figure}
\begin{center}
\includegraphics[width=0.45\textwidth]{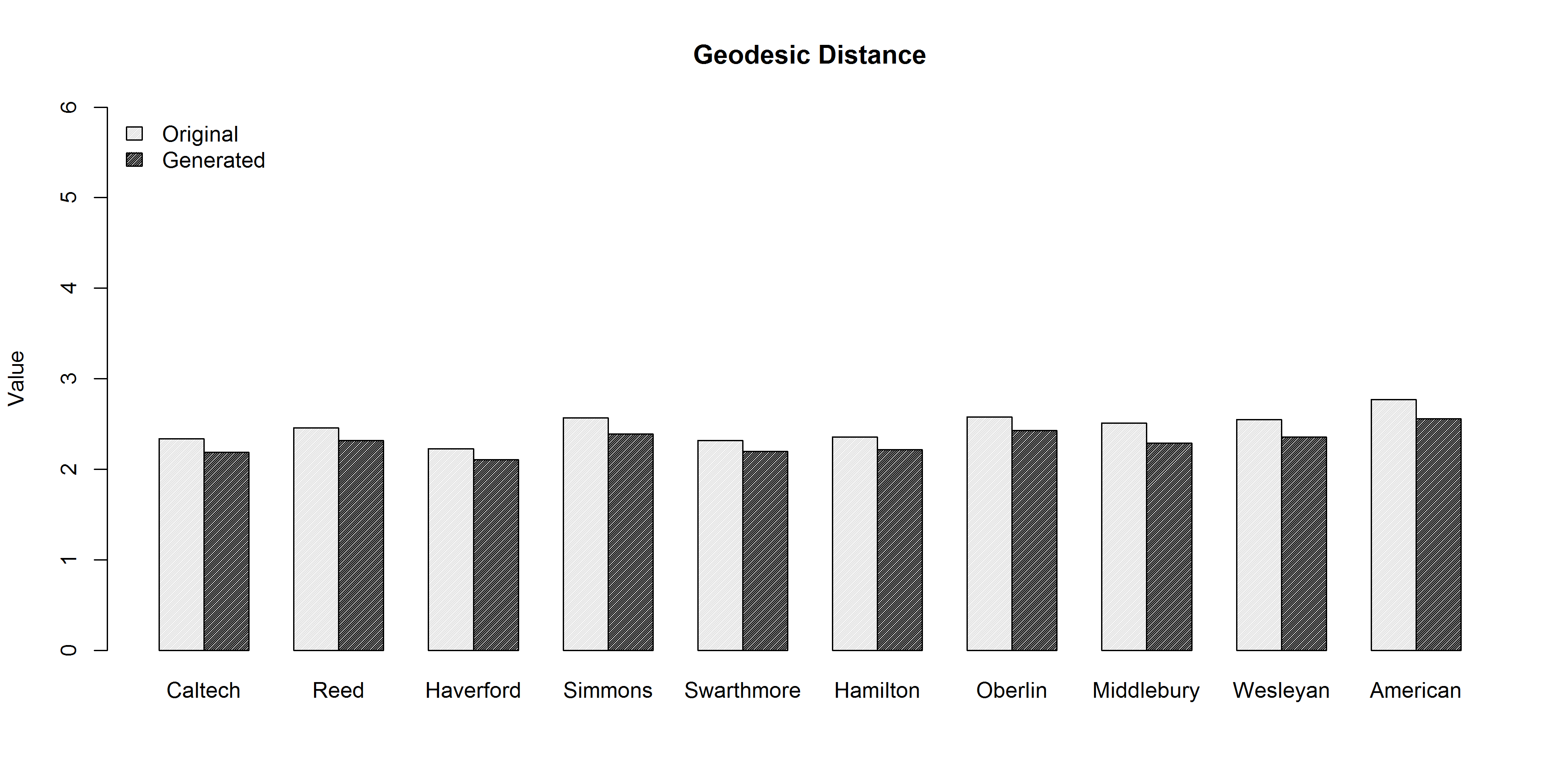}
\end{center}
\vspace{-10pt}
\caption{Comparative analysis of geodesic distances of the original graphs and the generated graphs.}
\label{fig::apl}
\end{figure}

\begin{figure}
\begin{center}
\includegraphics[width=0.45\textwidth]{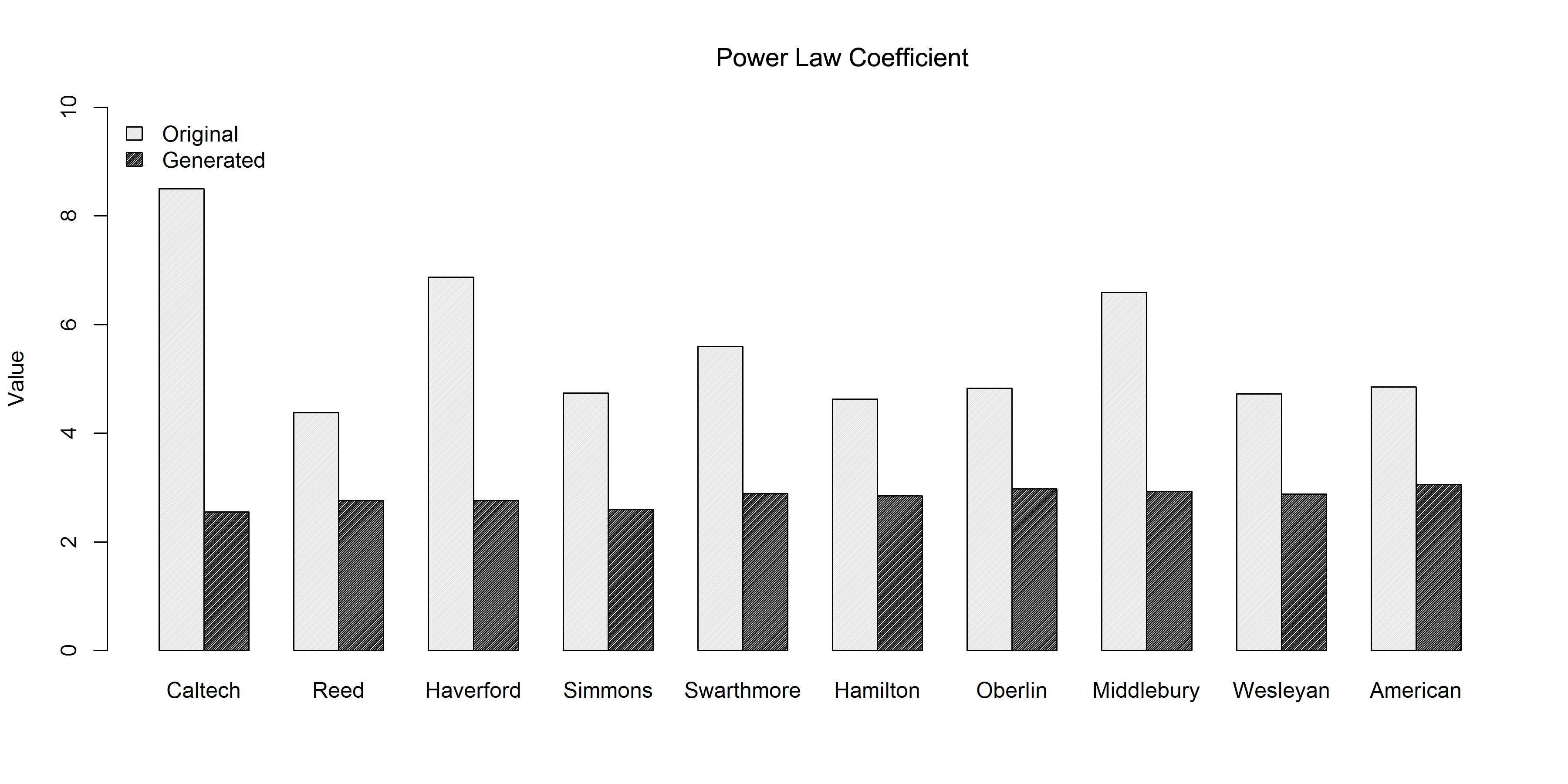}
\end{center}
\vspace{-10pt}
\caption{Comparative analysis of the power-law fit of the original graphs and the generated graphs.}
\label{fig::alpha}
\end{figure}

\begin{figure}
\begin{center}
\includegraphics[width=0.45\textwidth]{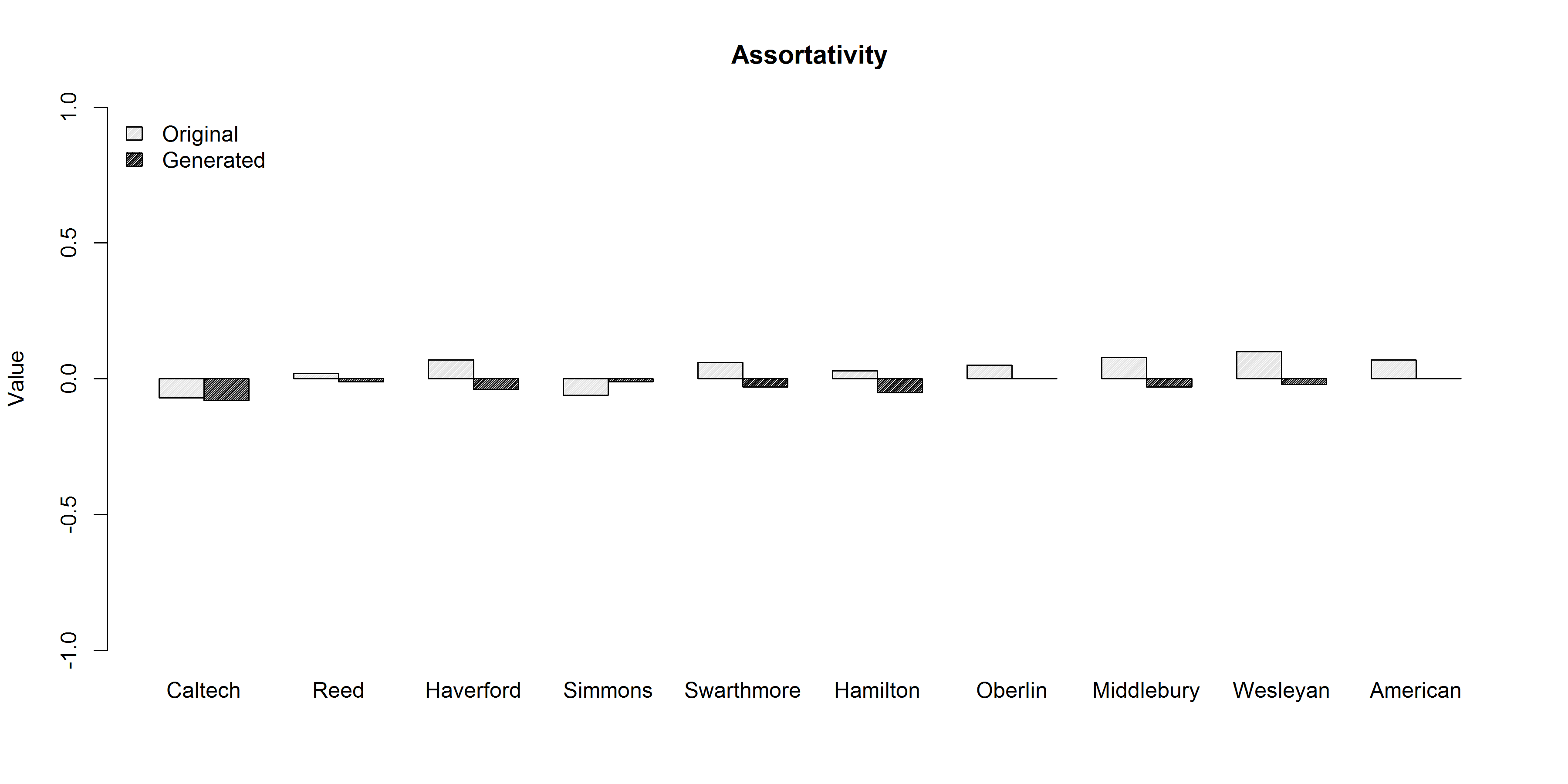}
\end{center}
\vspace{-10pt}
\caption{Comparative analysis of assortativity of the original graphs (empty circle) and the generated graphs (solid circle).}
\label{fig::assor}
\end{figure}

We compared the generated graphs using the proposed model with the original graphs using five metrics, the node-edge ratio often called density, the clustering coefficient, the average geodesic distance, the power-law fit (alpha) and assortativity. The results are shown in figure \ref{fig::density}, \ref{fig::cc}, \ref{fig::apl}, \ref{fig::alpha} and \ref{fig::assor}.

In case of density, the values generated by the proposed model are very similar to the original networks as shown in figure \ref{fig::density}. The proposed model uses the parameters $m_o$ and $m_f$ where the mean of the two approximately represents the average degree of nodes in the generated network. Increasing these values increase the overall density and vice versa. An important remark about $m_f$ is that this does not necessarily mean that the maximum degree of a node will not exceed $m_f$. These parameters signify the number of connections that a new entering node will form, not with whom they form so it is normal that due to preferential attachment, a new node might connect to a node with very high degree which might have connections more than $m_f$.

Figure \ref{fig::cc} shows the clustering coefficients of the original and the generated graphs. Again, we were able to generate values that are very close to the desired values. The clustering coefficient is controlled through the parameters $P(T_f)$, $P(T_L)$ and $L$ where $P(T_f)$ is the probability of triad formation taking place for the newly added node, $P(T_L)$ is the probability whether triad linkage will be performed among existing nodes and $L$ represents the number of such triads to be formed. High probability and increasing values of $L$ results in increasing the overall clustering coefficient of the generated network.

\begin{figure}
\begin{center}
\includegraphics[width=0.45\textwidth]{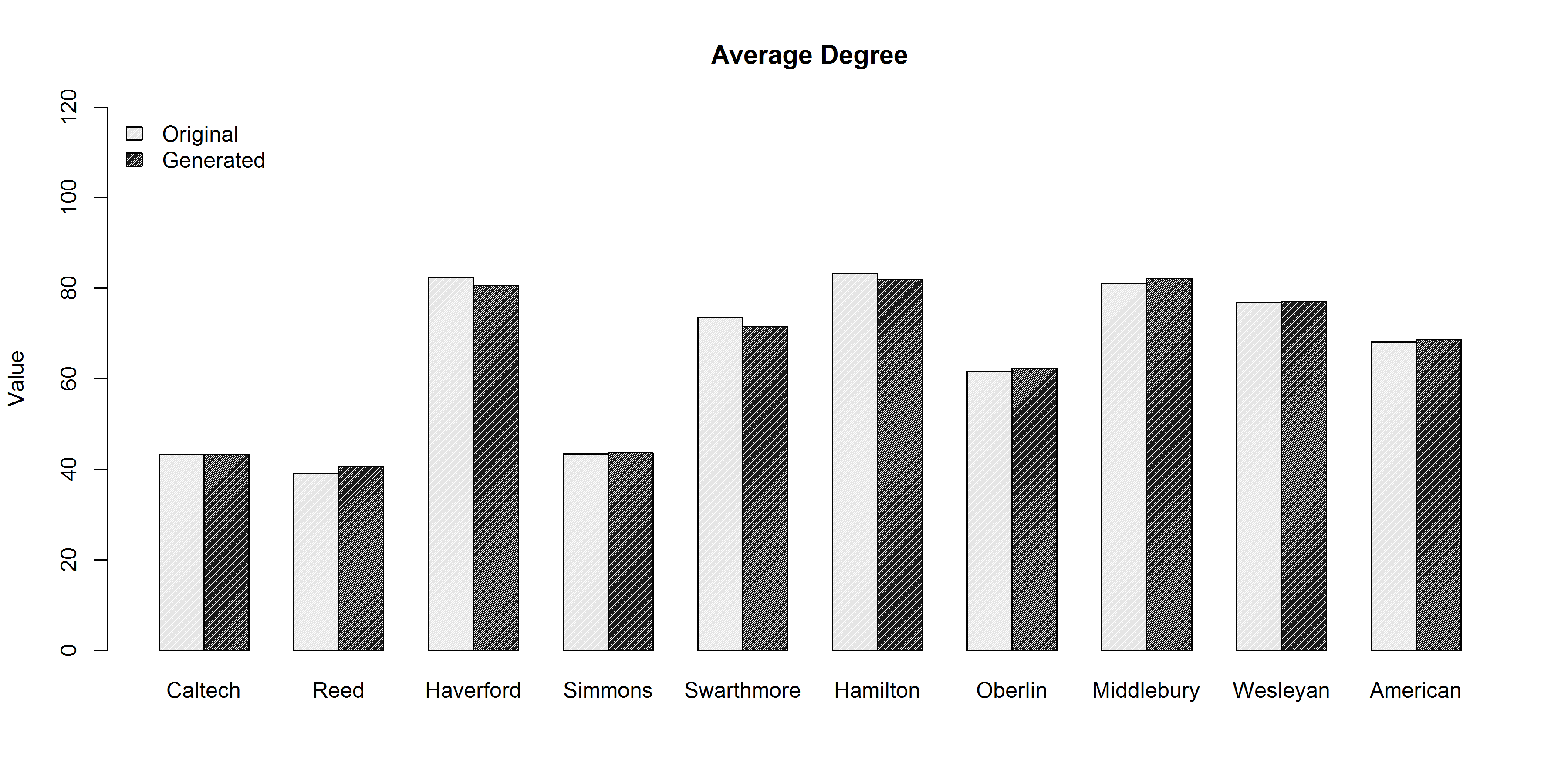}
\end{center}
\vspace{-10pt}
\caption{Comparative analysis of average degree of the original graphs and the generated graphs.}
\label{fig::avgdegree}
\end{figure}

In figure \ref{fig::apl}, we compare the geodesic distances of the networks again showing high similarity. We do not have any specific parameter to control this value but while calculating similarity based link formation, we consider preferential attachment based on degree connectivity, which results in both small geodesic distances for the generated graphs and their degree distribution following power-law as shown in figure \ref{fig::alpha}. All the generated networks have a power-law fit (alpha) between $1.9$ and $3.1$ suggesting scale free behavior of the proposed model. We were not able to match the power-law fit with that of the original facebook networks, since we incorporated the preferential attachment model \cite{barabasi99}, which is known to result in scale free degree distributions with power-law fit between $2$ or $3$. This fact is also well known for social networks but with the facebook datasets we used, the values of power-law fit are not between $2$ or $3$. Our experimentation suggests that we need to modify the existing methods to generate degree distributions to have a better fit rather than using the known preferential attachment model. One way to achieve a matching degree distribution is to use the model proposed by \cite{molloy95} which generates a network given a degree distribution. We did not use this method as it requires a static network whereas in our case, we propose a growing network generation model.

\begin{figure}
\begin{center}
\includegraphics[width=0.5\textwidth]{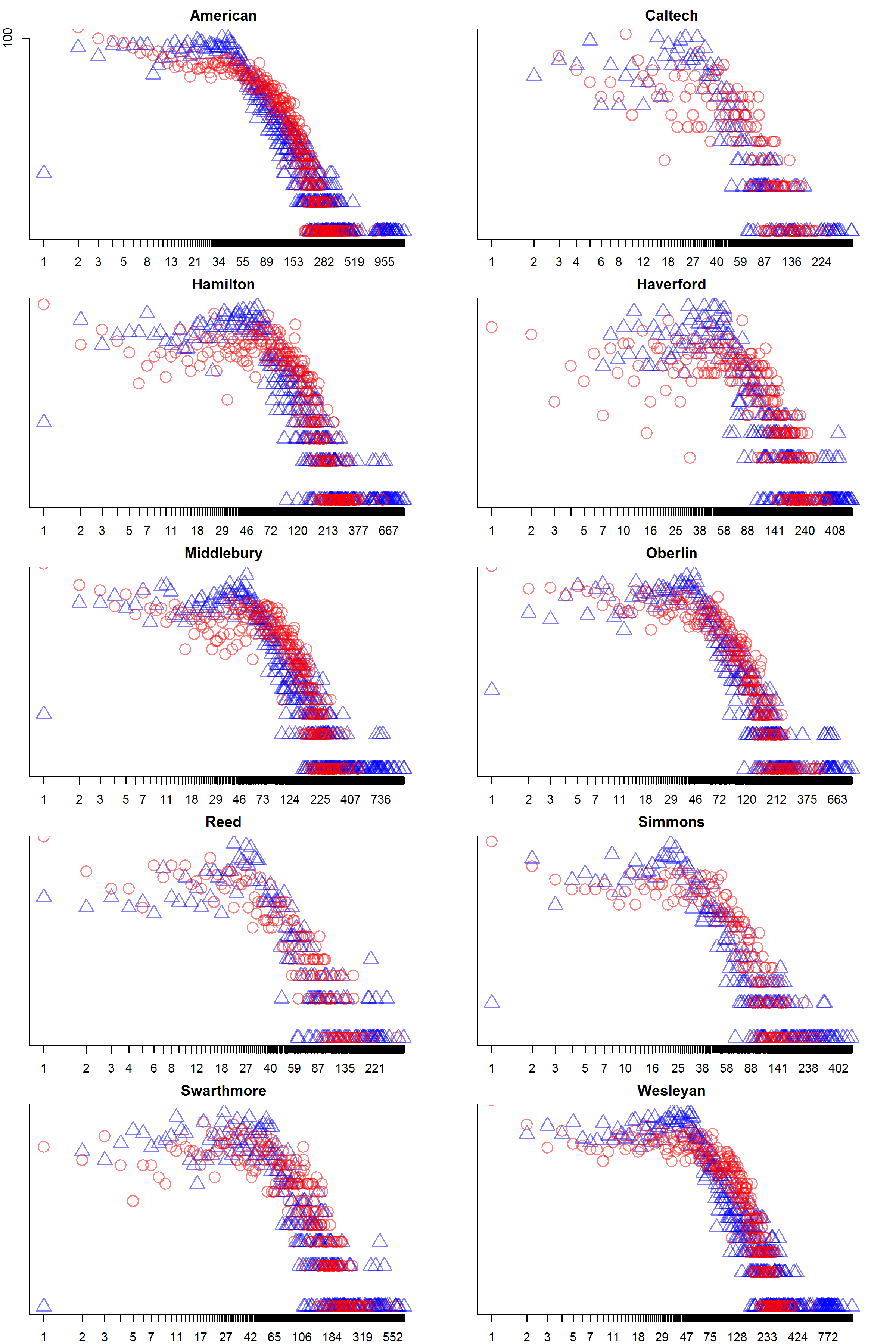}
\end{center}
\vspace{-10pt}
\caption{Comparative analysis of degree distribution of original and generated graphs for all datasets. Circles represents degree distribution of original graphs where triangles represents degree distribution of generated graphs.}
\label{fig::alldegdist}
\end{figure}

Figure \ref{fig::alldegdist} shows that although the power-law fit of the generated and original networks do not match, the general behavior of the degree distribution is very similar for all data sets except for Caltech and Haverford that have minor differences. The deviation in the power-law fit (figure \ref{fig::alpha}) are because of a few nodes in the original networks with extremely high variation in their frequency and connectivity, which in turn results in large values of alpha.

Figure \ref{fig::assor} shows the comparative assortativity values for the original and the generated networks. In case of Haverford, Swarthmore, Hamilton, and Middlebuy datasets, the original networks show a slightly positive assortativity, or assortative mixing, where as generated networks although have also very small values, they show negligible disassortativity. The differences between original and generated networks for all datasets are very small and all values are very close to $0$. These values are generated by setting the assortativity parameter $\gamma=1$ in the proposed algorithm. We study the effects of this parameter for different values on generated networks in section \ref{sec:empirical}.

The above comparative analysis shows high structural similarity among the original and generated networks for the ten facebook datasets considered in this paper. Table \ref{tbl:runs} shows the standard deviation for different metrics when calculated for $10$ runs for each facebook dataset using parameter values in table \ref{tbl:parameters}. Low standard deviation values for all metrics indicate the stability of the algorithm and lack significant structural variation in the generated networks.

\begin{table}[htbp]
  \centering
  \caption{Standard Deviation of network measures for multiple runs of the Proposed Model}
    \begin{tabular}{|r|r|r|r|r|r|r|}
    \hline
    \textbf{Dataset} & {\textbf{Density}} & {\textbf{Power Law}} & {\textbf{Geodesic}} & {\textbf{Clustering}} & {\textbf{Assortativity}} \bigstrut\\
         &  & {\textbf{Coefficient}} & {\textbf{Distance}} & {\textbf{Coefficient}} &  \bigstrut\\  
    
    \hline
    Caltech & 0.188 & 0.196 & 0.009 & 0.003 & 0.008 \bigstrut\\
    \hline
    Reed  & 0.039 & 0.114 & 0.012 & 0.004 & 0.012 \bigstrut\\
    \hline
    Haverford & 0.050 & 0.172 & 0.010 & 0.006 & 0.012  \bigstrut\\
    \hline
    Simmons & 0.031 & 0.245 & 0.009 & 0.006 & 0.008  \bigstrut\\
    \hline
    Swarthmore & 0.053 & 0.315 & 0.008 & 0.005 & 0.007  \bigstrut\\
    \hline
    Hamilton & 0.038 & 0.306 & 0.011 & 0.004 & 0.014 \bigstrut\\
    \hline
    Oberlin & 0.037 & 0.086 & 0.007 & 0.004 & 0.004  \bigstrut\\
    \hline
    Middlebury & 0.046 & 0.095 & 0.010 & 0.005 & 0.004  \bigstrut\\
    \hline
    Wesleyan & 0.015 & 0.101 & 0.005 & 0.004 & 0.005  \bigstrut\\
    \hline
    American & 0.022 & 0.115 & 0.008 & 0.002 & 0.005  \bigstrut\\
    \hline
    \end{tabular}%
   \label{tbl:runs}
\end{table}%

\section{Empirical Analysis of the Model Using Different Parameters}\label{sec:empirical}

In this section, we describe a number of experiments performed using the proposed model. The objective is to demonstrate the robustness, flexibility and control of the model as a function of different input parameters.

\begin{figure}
\begin{center}
\includegraphics[width=0.45\textwidth]{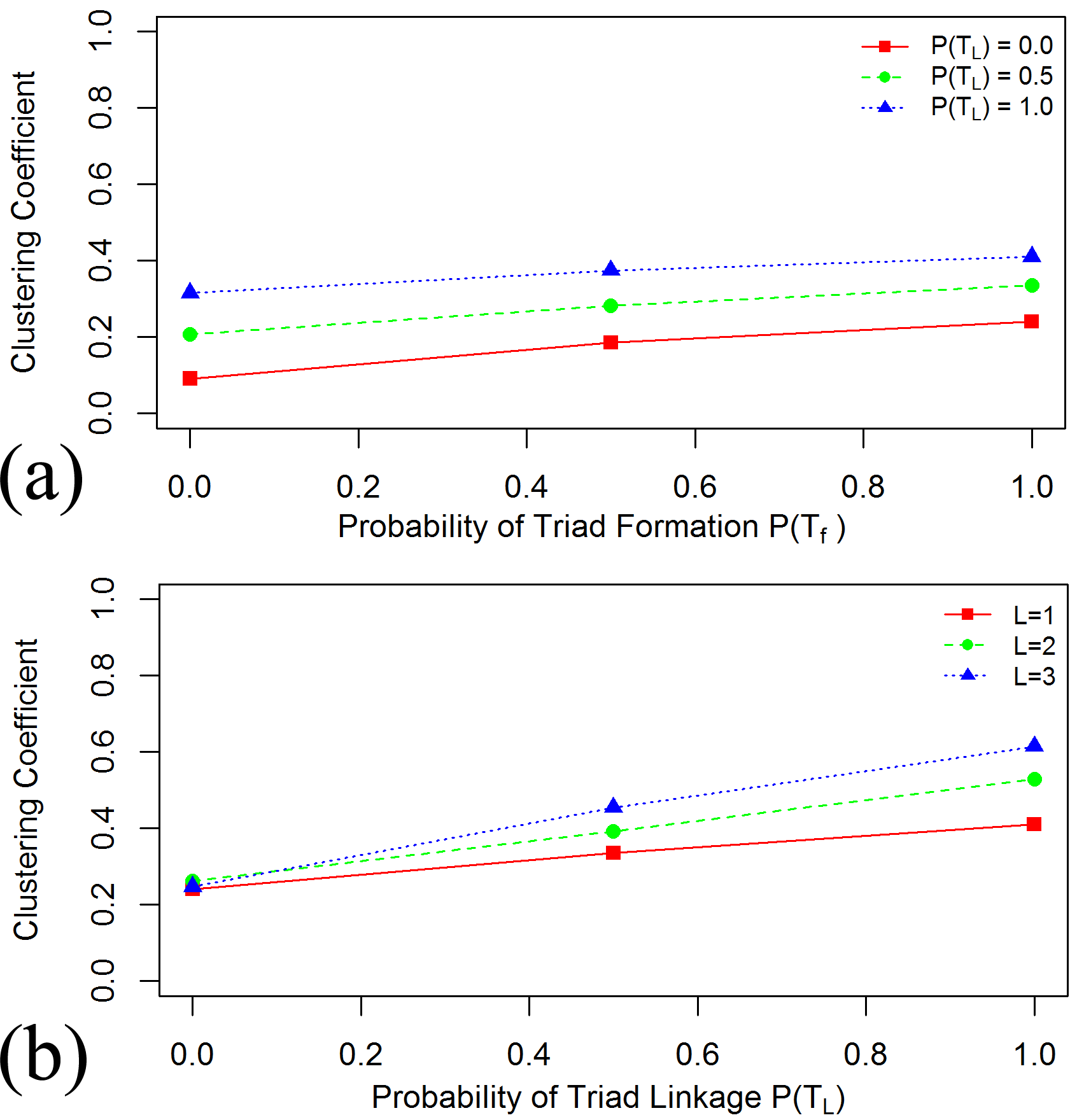}
\end{center}
\vspace{-10pt}
\caption{(a) Impact of Probability of Triad Formation $P(T_f)$ and Triad Linkage $P(T_L)$ on Clustering Coefficient. (b) Impact of Triad Linkage $P(T_L)$ and Triad Count $L$ on Clustering Coefficient.}
\label{fig::ccstudy}
\end{figure}

The first experiment studies the effects of controlling the overall clustering coefficient of the network. Figure \ref{fig::ccstudy} shows two graphs and the corresponding change in clustering coefficient as a function of parameters Triad Formation $P(T_f$), Triad Linkage $P(T_L)$ and Linkage Count $L$ changes the triad formation in the network generation process. Figure \ref{fig::ccstudy}(a) shows how Clustering coefficient of the overall graph increases with increasing values of $P(T_f$) and $P(T_L$) and figure \ref{fig::ccstudy}(b) shows the increasing effects of $P(T_L$) and $L$ on the generated networks. We explicitly separated the creation of triads in the network into two distinct steps using Triad Formation and Triad Linkage. This is to ensure two important societal features; Creation of new edges in existing nodes through triadic closures (which is not possible in many existing network models) and a new node only gaining limited links at the beginning (nodes select a few other nodes to connect) and then keep gaining links with the evolution of network.

%

\begin{figure}
\begin{center}
\includegraphics[width=0.5\textwidth]{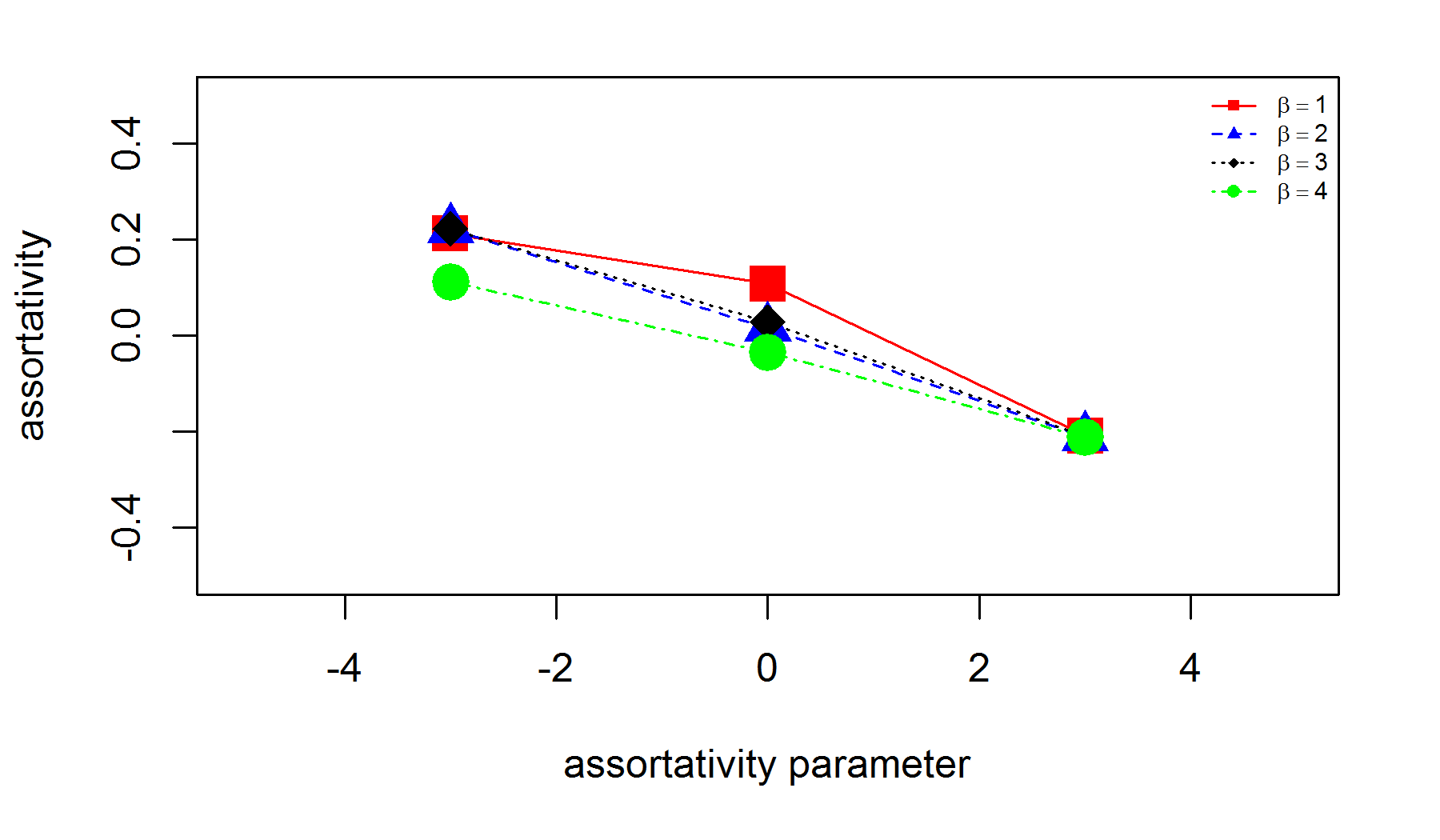}
\end{center}
\vspace{-10pt}
\caption{Impact of structural attributes $\beta$, and assortativity parameter $\gamma$ on assortative mixing based on node degree.}
\label{fig::assor-wd}
\end{figure}

The second experiment demonstrates the flexibility in generating assortative or disassortative networks. Recall from earlier sections that assortativity refers to property of nodes connecting to other nodes of similar node degree and is measured using the method proposed in \cite{newman03a}. Figure \ref{fig::assor-wd} shows how the assortativity is regulated using the  parameter $\gamma$ for different values of $\beta$, the weight associated to the structural characteristics in the network model. For negative values of $\gamma$ we obtain assortative networks with assortativity values greater than $0$ and for positive values of $\gamma$, we obtain disassortative networks with values less than $0$.
 
\begin{figure}
\begin{center}
\includegraphics[width=0.5\textwidth]{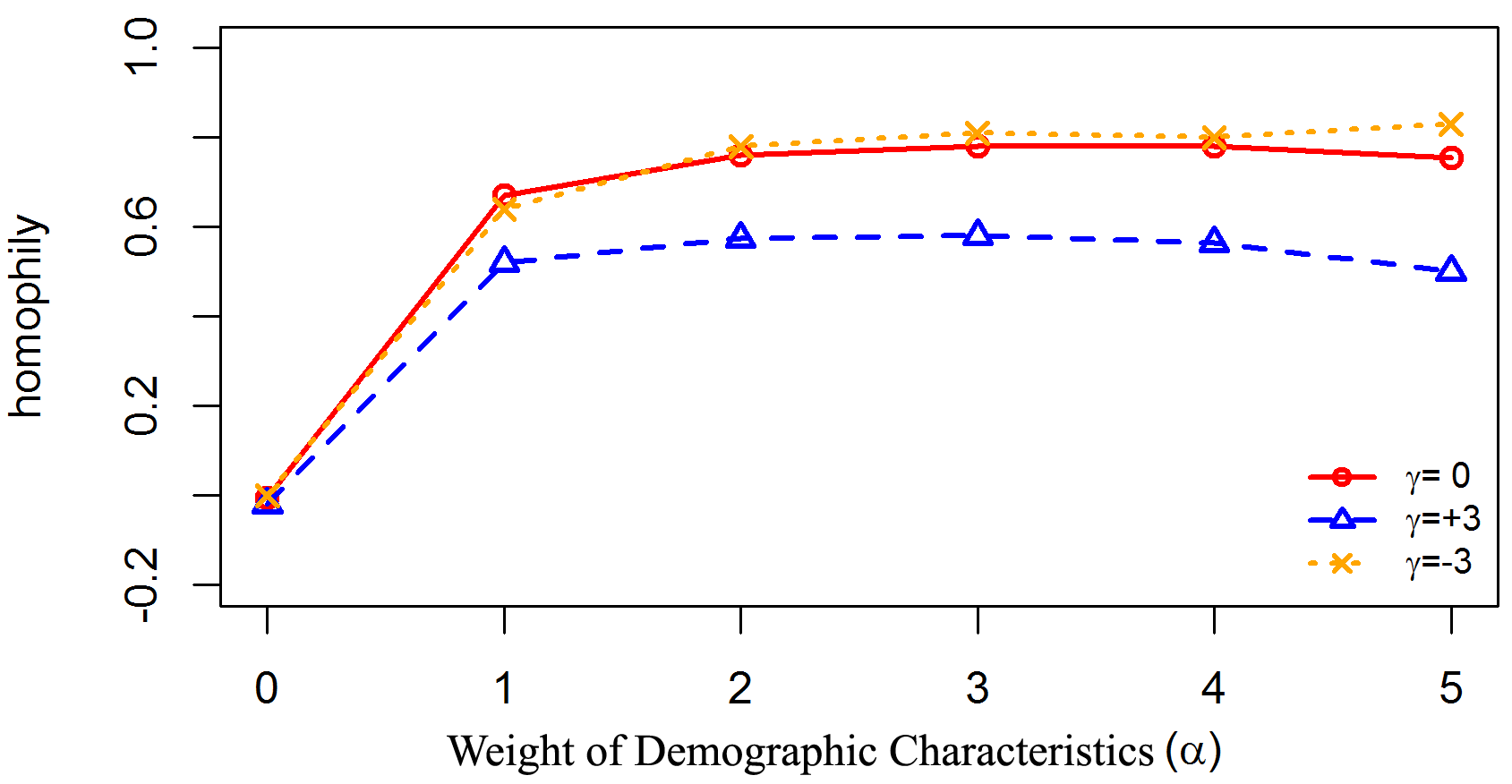}
\end{center}
\vspace{-10pt}
\caption{Impact of weight of demographic characteristics $\alpha$ and assortativity parameter $\gamma$ on demographic homophily of a sample generated networks.}
\label{fig::homophily}
\end{figure}

The last experiment studies the effect of parameters on demographic homophily of the generated networks. Figure \ref{fig::homophily} shows how the weight of the demographic characteristics play a role in the generation of homophilic networks. To calculate the homophily of networks, we again use the method proposed by \cite{newman03a} for numerical and categorical values assigned to nodes. We randomly assign five numerical values (1-5) and three categorical values (A,B,C) in equal proportion to generated networks of node size 1000. The experiment was repeated 10 times for each value to obtain average scores of homophily. High homophily values reflect high demographic similarity among neighboring nodes whereas low homophily demonstrates that more dissimilar nodes connect to each other in the network.

We tested the proposed model for different values of weight $\alpha$ which controls the importance of demographic characteristics. As $\alpha$ increases, homophily of the network increases and for low values of $\alpha$ we obtain networks with low demographic homophily. Another important aspect comes from the high values of $\gamma$ which controls the degree assortativity and enforces an increased importance for structural parameters. Thus $\gamma$ acts as a negative force to generate homophilic networks. As a result, for high values of $\beta$, we obtain networks with low homophily as shown in figure \ref{fig::homophily}.

In terms of time complexity, the algorithm grows quadratically with the size of the network. The slowest operation of the algorithm is the calculation of similarity in every iteration for each newly added node with all the existing nodes in the network. Since this similarity depends upon the number of nodal and structural attributes considered, the algorithm runs in $O(dn^{2})$ where $d$ is the number of attributes considered and $n$ is the number of nodes in the generated network.

\section{Conclusion}\label{sec:conclusion}

In this paper, we have proposed a network generation model based on demographic and structural characteristics in order to better understand and rationalize link formation among individuals. We used different Facebook datasets to validate our model as we successfully regenerated networks with the same densities, clustering coefficients, assortativity, degree distribution and geodesic distances. The model also allowed us to generate networks with degree assortativity and demographic homophily which are two important features of modern day networks. We demonstrated the effects of how structural and demographic properties can play a role to tune different structural properties of the generated networks.

Another important feature of social and other complex networks is the presences of community structures and we also foresee this amendment to the proposed model to generate more realistic networks. We also plan to extend this model for networks on interdependent multi-layered networks where each network layer can be generated from the proposed network model.

\bibliographystyle{IEEEtran}
\bibliography{visu}

\end{document}